\documentstyle{mn}

\newif\ifAMStwofonts

\newcommand{\etal}{{et al.}~}
\newcommand{\de}{\delta}
\newcommand{\te}{\theta}
\newcommand{\varte}{\vartheta}
\newcommand{\veps}{\varepsilon}
\newcommand{\Sig}{\Sigma}

\newcommand{\f}{\frac}
\newcommand{\s}{\sigma}
\newcommand{\bfx}{\bmath{x}}

\newcommand{\bfk}{\bmath{k}}
\newcommand{\bfv}{\bmath{v}}

\newcommand{\calD}{{\mathcal D}}

\newcommand{\calO}{{\mathcal O}}

\newcommand{\bc}{\begin{center}}
\newcommand{\be}{\begin{equation}}
\newcommand{\ee}{\end{equation}}
\newcommand{\ec}{\end{center}}
\newcommand{\lan}{\langle}
\newcommand{\ran}{\rangle}
\newcommand{\ov}{\overline}

\newcommand{\hmpc}{h^{-1}\,\rmn{Mpc}}
\newcommand{\kms}{{$\rmn{km}\; \rmn{s}^{-1}$}}

\ifoldfss
  \newcommand{\rmn}[1] {{\rm #1}}

  \ifCUPmtlplainloaded \else
    \NewTextAlphabet{textbfit} {cmbxti10} {}
    \NewTextAlphabet{textbfss} {cmssbx10} {}
    \NewMathAlphabet{mathbfit} {cmbxti10} {} 
    \NewMathAlphabet{mathbfss} {cmssbx10} {} 
  \fi
  \ifAMStwofonts
    \ifCUPmtlplainloaded \else
      \NewSymbolFont{upmath} {eurm10}
      \NewSymbolFont{AMSa} {msam10}
      \NewMathSymbol{\upi}     {0}{upmath}{19}
      \NewMathSymbol{\umu}     {0}{upmath}{16}
      \NewMathSymbol{\upartial}{0}{upmath}{40}
      \NewMathSymbol{\leqslant}{3}{AMSa}{36}
      \NewMathSymbol{\geqslant}{3}{AMSa}{3E}

       \let\le=\leqslant
       \let\ge=\geqslant
    \fi
  \fi
\fi 

\ifnfssone
  \newmathalphabet{\mathit}
  \addtoversion{normal}{\mathit}{cmr}{m}{it}
  \addtoversion{bold}{\mathit}{cmr}{bx}{it}
  \newcommand{\rmn}[1] {\mathrm{#1}}

  \newmathalphabet{\mathbfit} 
  \addtoversion{normal}{\mathbfit}{cmr}{bx}{it}
  \addtoversion{bold}{\mathbfit}{cmr}{bx}{it}
  \newmathalphabet{\mathbfss} 
  \addtoversion{normal}{\mathbfss}{cmss}{bx}{n}
  \addtoversion{bold}{\mathbfss}{cmss}{bx}{n}
  \ifAMStwofonts
    \ifCUPmtlplainloaded \else
      \UseAMStwoboldmath
      \makeatletter
      \new@mathgroup\upmath@group
      \define@mathgroup\mv@normal\upmath@group{eur}{m}{n}
      \define@mathgroup\mv@bold\upmath@group{eur}{b}{n}
      \edef\UPM{\hexnumber\upmath@group}
      \new@mathgroup\amsa@group
      \define@mathgroup\mv@normal\amsa@group{msa}{m}{n}
      \define@mathgroup\mv@bold\amsa@group{msa}{m}{n}
      \edef\AMSa{\hexnumber\amsa@group}
      \makeatother
      \mathchardef\upi="0\UPM19
      \mathchardef\umu="0\UPM16
      \mathchardef\upartial="0\UPM40
      \mathchardef\leqslant="3\AMSa36
      \mathchardef\geqslant="3\AMSa3E

       \let\le=\leqslant
       \let\ge=\geqslant
    \fi
  \fi
\fi 

\ifnfsstwo
  \newcommand{\rmn}[1] {\mathrm{#1}}

  \DeclareMathAlphabet{\mathbfit}{OT1}{cmr}{bx}{it}
  \SetMathAlphabet\mathbfit{bold}{OT1}{cmr}{bx}{it}
  \DeclareMathAlphabet{\mathbfss}{OT1}{cmss}{bx}{n}
  \SetMathAlphabet\mathbfss{bold}{OT1}{cmss}{bx}{n}
  \ifAMStwofonts
    \ifCUPmtlplainloaded \else
      \DeclareSymbolFont{UPM}{U}{eur}{m}{n}
      \SetSymbolFont{UPM}{bold}{U}{eur}{b}{n}
      \DeclareSymbolFont{AMSa}{U}{msa}{m}{n}
      \DeclareMathSymbol{\upi}{0}{UPM}{"19}
      \DeclareMathSymbol{\umu}{0}{UPM}{"16}
      \DeclareMathSymbol{\upartial}{0}{UPM}{"40}
      \DeclareMathSymbol{\leqslant}{3}{AMSa}{"36}
      \DeclareMathSymbol{\geqslant}{3}{AMSa}{"3E}

       \let\le=\leqslant
       \let\ge=\geqslant
    \fi
  \fi
\fi 

\ifCUPmtlplainloaded \else
  \ifAMStwofonts \else 
    \def\upi{\pi}
    \def\umu{\mu}
    \def\upartial{\partial}
  \fi
\fi

\title[Density from expansion and shear]{Large-scale density 
from velocity expansion and shear}
\author[M. J. Chodorowski]
       {Micha{\l} J. Chodorowski \\
 Copernicus Astronomical Center, Bartycka 18, 00--716 Warsaw, Poland 
}

\begin{document}

\maketitle

\begin{abstract}
I derive up to second order in Eulerian perturbation theory a new
relation between the weakly nonlinear density and velocity fields. In
the case of unsmoothed fields, density at a given point turns out to
be a purely local function of the expansion (divergence) and shear of
the velocity field. The relation depends on $\Omega$, strongly by the
factor $f(\Omega) \simeq \Omega^{0.6}$ and weakly by the factors
$K(\Omega) \propto \Omega^{-2/63}$ and $C(\Omega) \propto
\Omega^{-1/21}$. The Gramann solution is found to be equivalent to the
derived relation with the weak $\Omega$-dependence neglected. To make
the relation applicable to the real world, I extend it for the case of
smoothed fields. The resulting formula, when averaged over shear given
divergence, reproduces up to second order the density--velocity
divergence relation of Chodorowski \& {\L}okas; however, it has
smaller spread. It makes the formula a new attractive local estimator
of large-scale density from velocity.

\end{abstract}


\section{Introduction}

The value of $\Omega$ remains still one of the most intriguing
unknowns in cosmology today. The parameter $\Omega$, defined as the
ratio of the mean to the critical density is so crucial for cosmology
because its value determines the global geometry and ultimate fate of
Universe.

One way to measure $\Omega$ is to compare large-scale density fields
of galaxies with the corresponding fields of galaxy velocities.
Under widely accepted hypothesis of gravitational instability,
observed large-scale peculiar flows of galaxies (deviations from
Hubble flow) result from gravitational growth of initially small
cosmic mass fluctuations. The quantitative relation between the mass
density contrast field, $\de = \rho / \lan\rho\ran -1$, where
$\lan\rho\ran$ is the mean density, and the peculiar velocity field,
$\bfv$, can be deduced from the dynamical equations describing the
pressureless self-gravitating cosmic fluid. For small density
fluctuations linear theory can be applied. Linear theory predicts that
the density contrast is a linear and local function of the velocity
{\em divergence},
\be \de^{(1)}(\bfx) = -
f(\Omega)^{-1} \nabla \cdot \bfv^{(1)}(\bfx) \,.
\label{eq-1}
\ee

In the above, the function $f(\Omega) \simeq \Omega^{0.6}$ (see e.g.\
Peebles 1980) and the superscript `$(1)$' denotes the linear theory limit.  
(Distances are measured here in \kms, so the Hubble constant $H = 1$
in this system of units.) The above formula can be used to reconstruct
from a large-scale velocity field the linear mass density field, up to
an $\Omega$-dependent multiplicative factor $f(\Omega)$. The
comparison of the reconstructed mass field with the observed
large-scale galaxy density field may therefore serve as a method for
estimating $\Omega$ and as a test for the gravitational instability
hypothesis.

Indeed, a strong correlation between the galaxy density and velocity
divergence fields has been found in observations \cite{dbysd,hud,sig}.
However, equation~(\ref{eq-1}) assumes linear theory while the fields
in question are weakly nonlinear. Smoothing of the fields, necessary
among other things to reduce large individual distance-estimation
errors and the shot noise content, must be performed over a limited
scale in order to optimize the information present in the
finite-volume data. The {\sc potent} algorithm for the mass density
reconstruction from an observed radial velocity field currently
employs a Gaussian smoothing length of $1000$--$1200$ \kms
\cite{de,deket}. At these scales, typical (rms) galaxy density
fluctuations are of the order of several tens per cent, in
contradiction with an underlying assumption of equation~(\ref{eq-1})
that $\de \ll 1$. On the other hand, they are not in excess of unity,
wherefore the name ``weakly nonlinear''. The need for a weakly
nonlinear generalization of linear formula~(\ref{eq-1}) has been
quickly recognized. The present {\sc potent} algorithm uses the
formula of Nusser \etal (1991), which is the Zel'dovich (1970)
approximation expressed in Eulerian coordinates. However, N-body
simulations \cite{man,gan} have shown that though Zel'dovich
approximation does much better than linear theory
equation~(\ref{eq-1}), it still does not predict correctly the weakly
nonlinear density--velocity relation (hereafter DVR).

Weakly nonlinear regime is the regime of applicability of perturbation
theory. To begin with, linear theory solutions for the density and
velocity divergence fields that give rise to linear
equation~(\ref{eq-1}) are nothing but perturbative series truncated at
the lowest, i.e.\ first order terms. A natural way of extending linear
DVR into weakly nonlinear regime is thus to take into account higher
order perturbative contributions for density and velocity
divergence. This has been done by Chodorowski \& {\L}okas (1997a,
hereafter C{\L}97), who computed weakly nonlinear density--velocity
divergence relation up to third order in perturbation theory. (Second
order contributions were included already by Bernardeau 1992.) The
resulting extension of the linear formula offers also a method for
separating the effects of $\Omega$ and possible bias between galaxy
and mass distributions (C{\L}97; Bernardeau, Chodorowski \& {\L}okas 1997,
hereafter BC{\L}).

One might worry that the perturbative approximation to nonlinear DVR
breaks down soon after the linear relation does so. However, N-body
simulations \cite{bcl,gan,chod} show the opposite: the perturbative
formula is a very good robust fit to N-body results in the whole
cosmologically interesting range of smoothing radii.

Higher order perturbative solutions for density and velocity
divergence are nonlocal. As a result, the relation between weakly
nonlinear density and velocity divergence {\em at a given point\/} is
no longer deterministic. Still, since the spread comes exclusively
from higher order contributions the two fields remain strongly
correlated and the mean trend can serve as a useful local
approximation to the true nonlocal DVR. This is exactly what has been
calculated by C{\L}97, who found that the formula for the conditional mean
density given velocity divergence is given by the third-order
polynomial in velocity divergence. The reverse case (mean velocity
given density) has been calculated by Chodorowski \& {\L}okas
(1997b). Work is in progress on the theoretical prediction for the
spread \cite{clp}, as well as its measurement in N-body simulations
\cite{bcl,chod}.

Summarizing the above in the statistical language, the perturbative
polynomial in velocity divergence is an unbiased but non-zero variance
local estimator of density. It is then natural to ask a question:
among all unbiased local estimators of density from velocity, is it
the minimum-variance one? It is quite unlikely. The fact that in
weakly nonlinear regime density and velocity divergence at a given
point are not related in a deterministic way does not exclude a
possible existence of a purely local formula for density as a function
of some derivatives of the velocity field, $v_{i,j}$. The
irrotationality of the flow implies only that the tensor of velocity
derivatives is symmetric, $v_{i,j} = v_{j,i}$.  Hence, it has $6$
independent components, while the velocity divergence involves only
$3$ diagonal ones out of them.

Indeed, already mentioned Zel'dovich approximation which does involve
off-diagonal components is, like linear formula~(\ref{eq-1}),
deterministic. The formula based on Zel'dovich approximation is a
biased estimator, but it is not based on rigorous Eulerian
perturbation theory. The application of perturbation theory and the
inclusion of the off-diagonal components in the velocity derivatives
tensor may result in a local estimator which is both unbiased and has
smaller variance than the relation of C{\L}97. Why not simply
zero-variance?  To make the estimator applicable to the real world,
smoothing of the fields has to be included. Smoothing is a nonlocal
operation, hence it necessarily introduces some spread into any
nonlinear DVR; I will discuss this in more detail later on. Still,
since the density--velocity divergence relation is nonlocal already
for unsmoothed fields, for smoothed fields it can be expected to have
greater spread than the corresponding relation which is local when
unsmoothed.

This paper is devoted to constructing such a lower-variance estimator
of density from velocity. In section~2 I derive up to second order in
Eulerian perturbation theory a purely local relation between
unsmoothed density and velocity fields. In section~3 I relate it to
the Gramann~(1993) solution. Subsequently, in section~4 I derive from
it the unsmoothed density--velocity divergence relation. In section~5
I generalize my relation for the case of smoothed fields. Section~6 is
devoted to the comparison of the relation to N-body
simulations. Finally, I summarize the results in section~7.

\section{Derivation}

Let us expand the density contrast in a perturbative series, 
\be
\de = \de^{(1)} + \de^{(2)} + \de^{(3)} + \ldots \,.
\label{eq-2}
\ee 
In the above, $\de^{(p)}$ denotes the $p$-th order perturbative
contribution, which is of the order of $(\de^{(1)})^p$ \cite{fry,gor}.  
Introduce a variable proportional to the velocity divergence
\be 
\varte(\bfx, t)\equiv - f(\Omega)^{-1} \nabla_{\bfx}\cdot\bfv(\bfx,t)
\label{eq-3} 
\ee
and expand it as well, 
\be
\varte = \varte^{(1)} + \varte^{(2)} + \varte^{(3)} + \ldots \,.
\label{eq-4}
\ee

The linear theory solution, equation~(\ref{eq-1}), gives 
\be 
\de^{(1)}(\bfx) = \varte^{(1)}(\bfx) \,.
\label{eq-5}
\ee
From equations~(\ref{eq-2}), (\ref{eq-4}) and~(\ref{eq-5}) we have up
to second order 
\be 
\de(\bfx) = \varte(\bfx) + \de^{(2)}(\bfx) - \varte^{(2)}(\bfx) \,.
\label{eq-6}
\ee

The second order perturbative contributions for $\de$ and $\varte$ for
arbitrary $\Omega$ are \cite{bouch,bjdb}

\be 
\de^{(2)} = \f{1 + K}{2} (\de^{(1)})^{2} + \nabla_{\bfx} 
\de^{(1)} \cdot \nabla_{\bfx} \Phi^{(1)} + \f{1 - K}{2}
\Phi^{(1)}_{,ij} \Phi^{(1)}_{,ij}
\label{eq-7}
\ee
and 
\be 
\varte^{(2)} = C (\de^{(1)})^{2} + \nabla_{\bfx} \de^{(1)} 
\cdot \nabla_{\bfx} \Phi^{(1)} + (1 - C) \Phi^{(1)}_{,ij} 
\Phi^{(1)}_{,ij} \,.
\label{eq-8}
\ee
Here, $\Phi^{(1)}(\bfx, t)$ is the linear gravitational potential
satisfying the Poisson equation
\be
\Delta_{\bfx} \Phi^{(1)} = \de^{(1)} 
\label{eq-9} 
\ee
and I use the Einstein summation convention. The weakly
$\Omega$-dependent functions $K$ and $C$ are
\be
K(\Omega) = \f{3}{7} \Omega^{-2/63} 
\label{eq-10} 
\ee
and
\be
C(\Omega) = \f{3}{7} \Omega^{-1/21} \,.
\label{eq-11} 
\ee
The approximation~(\ref{eq-10}) is accurate to within $0.4$ per cent
in the range $0.05 < \Omega < 3$ \cite{bouch}
and~(\ref{eq-11}) to within $2$ per cent in the range $0.1 < \Omega <
10$ \cite{bjdb}. 

Subtracting equation~(\ref{eq-8}) from~(\ref{eq-7}) we have
\be
\de^{(2)} - \varte^{(2)} = 
\f{1 + K -2C}{2} (\varte^{(1)})^{2} - \f{1 + K -2C}{2}
\Phi^{(1)}_{,ij} \Phi^{(1)}_{,ij} 
\,.
\label{eq-13} 
\ee
Note that the terms $\nabla_{\bfx} \de^{(1)} \cdot \nabla_{\bfx}
\Phi^{(1)}$ have cancelled out. The well-known solution to
equation~(\ref{eq-9}) for the gravitational potential is
\be
\Phi^{(1)}(\bfx) = - \int \f{{\rm d}^3 x'}{4\pi} \f{\de^{(1)}(\bfx')}{\vert
\bfx - \bfx' \vert} \,.
\label{eq-12} 
\ee
Hence, the difference $\de^{(2)}(\bfx) - \varte^{(2)}(\bfx)$, in
addition to the local term $\sim (\varte^{(1)})^{2}(\bfx)$, contains a
nonlocal term due to the gravitational potential, $\sim \sum
\alpha_{\bfx' \bfx''}\de^{(1)}(\bfx') \de^{(1)}(\bfx'') \sim \sum
\alpha_{\bfx' \bfx''}\varte^{(1)}(\bfx') \varte^{(1)}(\bfx'')$. The
right-hand side of equation~(\ref{eq-13}) is thus not a function, but
a functional of $\varte$. On the one hand, this means that up to
second order in perturbation theory the velocity divergence field
still contains full information about the density field. On another
hand, however, this information is distributed non-locally, that is
the value of the density contrast at a given point $\bfx$ is only
determined by the values of the velocity divergence in all
space. Shortly, the relation between the weakly nonlinear density and
velocity divergence fields is deterministic, but nonlocal. Such a
relation is of no use for cosmology, since we have at our disposal
only surveys of limited volume, with uncomplete sky coverage, uneven
sampling, etc. The key point is to find a local relation between the
density and velocity fields.

One of the ways to do this is to approximate the nonlocal-in-$\varte$
term in equation~(\ref{eq-13}) by its conditional mean, given the
velocity divergence at a point $\bfx$ \cite{b92,cl97}. The resulting
density--velocity divergence relation is then local by construction
but clearly not deterministic; in other words it has some spread. The
idea of the present paper is different: to recast the
nonlocal-in-$\varte$ term to the form involving other local
derivatives of the velocity field.

Since the velocity field in weakly nonlinear regime remains
irrotational it can be expressed as a gradient of some velocity
potential $\Phi_{\bfv}$,
\be
\bfv = - \nabla_{\bfx} \Phi_{\bfv} \,.
\label{eq-14} 
\ee
Taking the divergence of the above equation and using
equation~(\ref{eq-1}) shows that in linear regime the velocity
potential is proportional to the gravitational potential,
\be
\Phi_{\bfv}^{(1)} =  f(\Omega) \Phi^{(1)} \,,
\label{eq-15} 
\ee
and equations~(\ref{eq-14})--(\ref{eq-15}) yield
\be 
\Phi^{(1)}_{,ij} = - f^{-1} v^{(1)}_{i,j} \,.
\label{eq-16} 
\ee

Let us decompose the symmetric tensor of velocity derivatives into its
trace (expansion), $\te$, and the traceless part (shear), $\s_{ij}$,
\be
v_{i,j} = \f{1}{3} \te \de_{ij} + \s_{ij}
\,,
\label{eq-17} 
\ee
where in general
\be
\s_{ij} \equiv \f{1}{2} \left( v_{i,j} +
v_{j,i} \right) - \f{1}{3} \te \de_{ij}
\label{eq-18} 
\ee
and
\be
\te \equiv \nabla_{\bfx} \cdot \bfv = v_{k,k}
\,.
\label{eq-19} 
\ee
The symbol $\de_{ij}$ denotes the Kronecker's delta. Note by comparing 
with equation~(\ref{eq-3}) that 
\be
\varte = - f^{-1} \te
\,.
\label{eq-21} 
\ee

We have
\be
v_{i,j} v_{i,j} = \f{1}{3} \te^2 + \s^2
\,,
\label{eq-22} 
\ee
where $\s^2$ is the shear scalar
\be
\s^2 = \s_{ij} \s_{ij} \,.
\label{eq-23} 
\ee
From equations~(\ref{eq-16}) and~(\ref{eq-22}) taken at linear order
we obtain 
\be
\Phi^{(1)}_{,ij} \Phi^{(1)}_{,ij} = \f{1}{3} f^{-2} (\te^{(1)} )^2 
+ f^{-2} (\s^{(1)})^2
\,.
\label{eq-24} 
\ee
Introducing the above equation to equation~(\ref{eq-13}) and
using~(\ref{eq-21}) yields
\begin{eqnarray}
\de^{(2)} - \varte^{(2)} 
&=& \f{1 + K -2C}{3} f^{-2} (\te^{(1)})^{2} \nonumber \\
&-& \f{1 + K -2C}{2} f^{-2} (\s^{(1)})^2
\,.
\label{eq-25} 
\end{eqnarray}

Denote the rms fluctuation of the linear density field by $\veps$,
$\veps^2 = \lan (\de^{(1)})^2 \ran$. Perturbation theory predicts
that $\te^{(p)} \sim \de^{(p)} \sim \veps^p$ so we have $\te^2 =
(\te^{(1)} + \te^{(2)} + \ldots)^2 = (\te^{(1)})^2 + \calO(\veps^3)$
and similarly for $\s^2$. Therefore, constructing weakly nonlinear
density--velocity relation up to terms quadratic in $\veps$ we can
substitute $(\te^{(1)})^2$ by $\te^2$ and $(\s^{(1)})^2$ by
$\s^2$. By doing so and introducing equation~(\ref{eq-25})
into~(\ref{eq-6}) we finally obtain

\begin{eqnarray} 
\de(\bfx) 
&=& - f^{-1} \te(\bfx) + \f{1 + K -2C}{3} f^{-2} 
\te^2(\bfx)  \nonumber \\
&-& \f{1 + K -2C}{2} f^{-2} \s^2(\bfx) + \calO(\veps^3)
\,.
\label{eq-26} 
\end{eqnarray}

This is the DVR computed up to second order in perturbation
theory. The density is a local function of the expansion and the shear
of the velocity field. A similar relation has been independently
derived by Catelan \etal (1995), though without the decomposition of the
term $v_{i,j} v_{i,j}$ into the expansion and the shear (cf.\ eq.~[41]
of Catelan \etal 1995). The relation depends on $\Omega$, strongly by
the factor $f$ and weakly by the factors $K$ and $C$ given by
equations~(\ref{eq-10})-(\ref{eq-11}). Neglecting the weak
$\Omega$-dependence we obtain

\be \de(\bfx) = - f^{-1} \te(\bfx) +
\f{4}{21} f^{-2} \te^2(\bfx) - \f{2}{7} f^{-2} \s^2(\bfx) \,.
\label{eq-28} 
\ee

Formula~(\ref{eq-26}) for density involves only first derivatives of
the velocity field. We owe this to the fortunate cancellation of the
terms $\nabla_{\bfx} \de^{(1)} \cdot \nabla_{\bfx} \Phi^{(1)}$ in
equation~(\ref{eq-13}). Had not these terms cancelled, we would have
had on the right hand side of equation~(\ref{eq-26}) an extra
contribution $\sim \nabla_{\bfx}\Phi^{(1)} \cdot \nabla_{\bfx}
\te^{(1)} \sim v_{j} v_{i,ij} + \calO(\veps^3)$, involving the second
derivatives. The practical applicability of such a formula to
extracting density from a very noisy velocity field would be doubtful.

\section{Relation to the Gramann solution}

Gramann (1993) derived DVR up to second order in Lagrangian
perturbation theory. The form of this relation expressed in Eulerian
coordinates is 

\be
\de(\bfx) = - f^{-1} \te(\bfx) + \f{4}{7} f^{-2} m_{\bfv}(\bfx)
\,,
\label{eq-29} 
\ee
where
\be
m_{\bfv} = \sum_{i<j}^3 \left( v_{i,i} v_{j,j} - v_{i,j} v_{j,i}
\right) 
\,,
\label{eq-30} 
\ee
or, explicitly
\begin{eqnarray}
m_{\bfv} 
&=& v_{1,1} v_{2,2} + v_{1,1} v_{3,3} + v_{2,2} v_{3,3}
\nonumber \\
&-& v_{1,2} v_{2,1} - v_{1,3} v_{3,1} - v_{2,3} v_{3,2}
\,.
\label{eq-31} 
\end{eqnarray}

In the following I will relate equation~(\ref{eq-29}) to my solution
for weakly nonlinear DVR calculated in the previous section. Let us
decompose the tensor of velocity derivatives more generally into the
expansion, shear, and vorticity (the asymmetric part), respectively,
\be
v_{i,j} = \f{1}{3} \te \de_{ij} + \s_{ij} - \f{1}{2} 
\epsilon_{ijk} \omega_{k}
\,.
\label{eq-32} 
\ee
Here, $\epsilon_{ijk}$ is the completely antisymmetric tensor;
$\epsilon_{123}= 1$. After applying the above decomposition, a few
lines of algebra yield
\begin{eqnarray}
v_{1,1} v_{2,2} + v_{1,1} v_{3,3} + v_{2,2} v_{3,3} 
&=& \s_{11} \s_{22} + \s_{11} \s_{33} + \s_{22} \s_{33}
\nonumber \\
&+& \f{1}{3} \te^2
\,.
\label{eq-33} 
\end{eqnarray}
Similarly,
\be
v_{1,2} v_{2,1} + v_{1,3} v_{3,1} + v_{2,3} v_{3,2} = 
\s_{12}^{2} + \s_{13}^{2} +  \s_{23}^{2} - 
\f{1}{4} \omega^2
\,,
\label{eq-34} 
\ee
where $\omega^2$ is the vorticity scalar, 
\be
\omega^2 \equiv \bmath{\omega} \cdot \bmath{\omega} = 
\omega_{k} \omega_{k}
\,.
\label{eq-35} 
\ee

By definition,
\be
\s^2 = \s_{ij} \s_{ij} = 
2 \left( \s_{12}^{2} + \s_{13}^{2} +  \s_{23}^{2}
\right) +  \s_{11}^{2} + \s_{22}^{2} +  \s_{33}^{2} 
\,.
\label{eq-36} 
\ee
Using the above equation and the identity
\be
\left( \s_{11} + \s_{22} +  \s_{33} \right)^2 = 0 \,,
\label{eq-37} 
\ee
we can recast equation~(\ref{eq-34}) to the form
\begin{eqnarray}
v_{1,2} v_{2,1} + v_{1,3} v_{3,1} + v_{2,3} v_{3,2} 
&=& \s_{11} \s_{22} + \s_{11} \s_{33} + \s_{22} \s_{33} 
\nonumber \\
&+& \f{1}{2} \s^2 - \f{1}{4} \omega^2
\,.
\label{eq-38} 
\end{eqnarray}

Using equations~(\ref{eq-33}), (\ref{eq-38}) and~(\ref{eq-31})
we obtain
\be
m_{\bfv} = \f{1}{3} \te^2 - \f{1}{2} \s^2 + \f{1}{4}
\omega^2 
\,.
\label{eq-39} 
\ee
The final step is to use this result in equation~(\ref{eq-29}),
the result is
\begin{eqnarray}
\de(\bfx) 
&=& - f^{-1} \te(\bfx) + \f{4}{21} f^{-2} 
\te^2(\bfx) - \f{2}{7} f^{-2} \s^2(\bfx) 
\nonumber \\
&+& \f{1}{7} f^{-2} \omega^2(\bfx)
\,.
\label{eq-40} 
\end{eqnarray}

Thus, the weakly nonlinear density in a given point is in general
determined by the local values of the three scalars that can be
constructed from the derivatives of the velocity field: the expansion,
shear and vorticity. Since before shell crossing a cosmic velocity
field is irrotational we can in the above equation drop out the
vorticity term. Then the equation coincides exactly with
equation~(\ref{eq-28}) of the previous section. Equation~(\ref{eq-28})
is the DVR up to second order in perturbation theory with the weak
$\Omega$-dependence neglected. Summarizing, the Gramann (1993)
solution~(\ref{eq-29}) is the second-order DVR with only the strong
$\Omega$-dependence included.

Equation~(\ref{eq-40}) bears some resemblance to the
Raychaudhuri~(1955) differential equation for the evolution of the
velocity expansion. There, the source terms are similarly proportional
to density, expansion, shear and vorticity (cf.~eq.~[22.14] of Peebles
1980).

\section{Density---velocity divergence relation}

As already mentioned in section~2, density and velocity divergence at
a given point are not related in a deterministic way. Let us rewrite
equation~(\ref{eq-26}) in the form
\begin{eqnarray} 
\de(\bfx) 
&=& \varte(\bfx) + \f{1 + K -2C}{3} \varte^2(\bfx) - 
\f{1 + K -2C}{2} \Sig^2(\bfx) 
\nonumber \\
&+& \calO(\veps^3)
\,,
\label{eq-41} 
\end{eqnarray}
where $\varte$ is related to $\te$ by equation~(\ref{eq-21}) and I 
have defined 
\be
\Sig_{ij} \equiv - f^{-1} \s_{ij} 
\,.
\label{eq-42} 
\ee
The spread in the $\de$--$\varte$ relation clearly comes from the
shear. The mean trend, defined as mean $\de$ given $\varte$, is
\be
\lan \de \ran|_{\varte} = \varte + \f{1 + K -2C}{3} 
\varte^2 - \f{1 + K -2C}{2} \lan \Sig^2 \ran|_{\varte} 
+ \calO(\veps^3)
\,.
\label{eq-43} 
\ee
By definition of the conditional moment, 
\be
\lan \Sig^2 \ran|_{\varte} = \f{\int \Sig^2 \, p(\varte,\Sig)\, 
{\rm d}\Sig}{p(\varte)} 
\,,
\label{eq-44}
\ee
where $p(\varte,\Sig)$ is the joint probability distribution function
(PDF) for expansion and the shear scalar. It is sufficient to know the
form of this PDF for linear $\varte$ and $\Sig$ since already at the
lowest order $\Sig^2 \sim \veps^2$ and nonlinear corrections are $\sim
\calO(\veps^3)$.

How to derive the joint distribution for $\varte^{(1)}$ and
$\Sig^{(1)}$? In the derivation of its general properties I will
follow Juszkiewicz \etal (1995; Appendix A). I assume that the initial
conditions are Gaussian. Under this assumption, both $\varte^{(1)}$
and five independent components of the shear tensor $\Sig_{ij}^{(1)}$
are Gaussian distributed. Consequently,
$p(\varte^{(1)},\Sig_{ij}^{(1)})$ is a multivariate Gaussian, entirely
determined by its covariance matrix. It is more convenient to compute
the coefficients of this matrix in the Fourier space. The Fourier
transform of $\varte^{(1)}$ is obviously equal to the Fourier
transform of the linear density field, $\varte^{(1)}_{\bfk} =
\de^{(1)}_{\bfk}$. Thereafter I will drop out the superscripts
`${(1)}$'. The power spectrum $P(k)$ is defined by the relation

\be
\lan \de_{\bfk} \de_{\bfk'} \ran = (2 \pi)^3 \de_D(\bfk + \bfk') P(k)
\,,
\label{eq-45} 
\ee
where $\de_D$ denotes the Dirac's delta. The Fourier transform of a
shear component is
\be
\left( \Sig_{ij} \right)_{\bfk} = \left( \hat{k_i} \hat{k_j} -
\f{1}{3} \de_{ij} \right) \de_{\bfk} 
\,,
\label{eq-46} 
\ee
where $\hat{k_i} \equiv k_i/k$. We have
\begin{eqnarray}
\lan \varte \Sig_{ij} \ran 
&=& \int \f{{\rm d}^3 k \, {\rm d}^3 k'}{(2 \pi)^6} \, 
\left( \hat{k_i}' \hat{k_j}' - \f{1}{3} \de_{ij} \right)
\lan \de_{\bfk} \de_{\bfk'} \ran 
{\rm e}^{- i (\bfk + \bfk') \cdot \bfx}
\nonumber \\
&=& \int \f{{\rm d}^3 k}{(2 \pi)^3} \, 
\left( \hat{k_i} \hat{k_j} - \f{1}{3} \de_{ij} \right)
P(k)
\nonumber \\
&=& 0
\,.
\label{eq-47} 
\end{eqnarray}
The second step uses equation~(\ref{eq-45}) and the last one is
obvious by symmetry. It means that the linear shear components are
uncorrelated with the linear velocity divergence. In general, the
uncorrelation of random variables is only a necessary condition for
their statistical independence; for Gaussian variables however it is
also the sufficient one. Therefore, $\varte$ and $\Sig_{ij}$ are
statistically independent; consequently the variables $\varte$ and
$\Sig = (\Sig_{ij}\Sig_{ij})^{1/2}$ are also statistically
independent. It implies
\be 
p(\varte,\Sig) = p(\varte)\, p(\Sig)
\,.
\label{eq-48} 
\ee

Using this result in equation~(\ref{eq-44}) yields
\be
\lan \Sig^2 \ran|_{\varte} = \int \Sig^2 p(\Sig) \, {\rm d}\Sig 
\,,
\label{eq-49}
\ee
that is the mean value of $\Sig^2$ does not depend on $\varte$ and is
equal to the ordinary mean. We have
\be
\lan \Sig_{ij} \Sig_{ij} \ran = \lan \Phi_{,ij} \Phi_{,ij} \ran - 
\f{1}{3} \lan \de^2 \ran 
\,.
\label{eq-50}
\ee
The integration by parts of the term $\lan \Phi_{,ij} \Phi_{,ij} \ran$
yields $\lan \Phi_{,ii} \Phi_{,jj} \ran = \lan \de^2 \ran = \veps^2$,
hence 
\be
\lan \Sig^2 \ran = \f{2}{3} \veps^2
\,.
\label{eq-51}
\ee
(cf.\ Silk 1974). Introducing this in equation~(\ref{eq-43}) we 
finally obtain
\be
\lan \de \ran|_{\varte} = \varte + a_2 ( \varte^2 - \veps^2 ) 
+ \calO(\veps^3)
\,,
\label{eq-52} 
\ee
where
\be
a_2 = \f{1 + K -2C}{3}  
\,.
\label{eq-53} 
\ee

Equations~(\ref{eq-52})-(\ref{eq-53}) agree with the results of C{\L}97
obtained by a completely different method, namely the Edgeworth 
expansion of the joint PDF for the variables $\de$ and $\varte$. 
Specifically, formula~(\ref{eq-53}) for the coefficient $a_2$
coincides with equation~(81) of C{\L}97 (for the case of no smoothing).

I will now compute the spread around the mean trend, or the
conditional variance. Hereafter in this section I will neglect the
weak dependence of equations~(\ref{eq-41}) and~(\ref{eq-43}) on
$\Omega$. We then have
\be
\left.\left\lan \left( \de - \lan \de \ran|_{\varte} \right)^2 
\right\ran\right|_{\varte} = \f{4}{49}
\left.\left\lan \left( \Sig^2 - \lan \Sig^2 \ran|_{\varte} \right)^2 
\right\ran\right|_{\varte}
+ \calO\left(\veps^6\right)
\,.
\label{eq-54} 
\ee
It is known that $\Sig^2$ is $\chi^2$-distributed with 5 degrees of
freedom; the variance of the Gaussian variable underlying the
distribution is $2/15 \veps^2$ (G{\'o}rski 1988; cf.\ also Groth,
Juszkiewicz \& Ostriker 1989). Therefore
\be 
\left.\left\lan \left(
\Sig^2 - \lan \Sig^2 \ran|_{\varte} \right)^2
\right\ran\right|_{\varte} = 2 \cdot 5 \cdot \f{4}{225} \veps^4 =
\f{8}{45} \veps^4 \,,
\label{eq-55} 
\ee
hence finally
\be
\left.\left\lan \left( \de - \lan \de \ran|_{\varte} \right)^2 
\right\ran\right|_{\varte} = \f{32}{2205} \veps^4 + 
\calO\left(\veps^6\right)
\,.
\label{eq-56} 
\ee

The above formula exactly coincides with the result of Chodorowski,
{\L}okas \& Pollo (1997; for the case of no smoothing), obtained by
means of the Edgeworth expansion.

And if initial conditions are non-Gaussian? In this case
property~(\ref{eq-47}) still holds as its derivation does not require
any assumption about initial conditions. Therefore, $\varte$ and
$\Sig_{ij}$ remain uncorrelated. It however does not necessarily mean
that they are statistically independent. Uncorrelated non-Gaussian
variables may be, but equally well may not be, statistically
independent; see Kendall \& Stuart (1973) for the examples of both
cases. Catelan \etal (1997) show that for a certain class of
non-Gaussian models $\varte$ and $\Sig$ are indeed independent. For
this class of non-Gaussian models the density--velocity divergence
relation is therefore the same as for Gaussian initial conditions,
equation~(\ref{eq-52}). Only the spread around the mean trend is
different since $\Sig^2$ is no longer $\chi^2$-distributed. The form
of the $\de$-$\varte$ relation for other non-Gaussian models remains
to be investigated.

\section{Effects of smoothing}

Ganon \etal (1997) test various approximations for weakly nonlinear
DVR by the means of N-body simulations. Among the approximations
considered is the Gramann (1993) solution. Since the Gramann solution
is equivalent to equation~(\ref{eq-26}) with the weak
$\Omega$-dependence neglected (section~3), its properties essentially
reflect the properties of the present solution. Ganon \etal (1997)
plot the difference between the approximate and true density as a
function of the true density, $\calD = \calD(\de)$. The Gramann
solution gives significant residuals which have a parabolic form. Is
equation~(\ref{eq-26}) thus incorrect?  There is certainly no error in
its derivation, but it cannot be straightforwardly applied to the {\em
smoothed} density and velocity fields that were estimated from N-body
by Ganon \etal (1997).

Inferring the fields from observations one has to introduce smoothing.
The first reason for doing so is to reduce the effects of large
individual distance-estimation errors and the shot noise. The second
one is that only the field of radial peculiar velocities is directly
measured; to reconstruct from it the full three-dimensional velocity
field we need to assume its potentiality. This assumption is valid
only before trajectory crossing (where the Kelvin's circulation
theorem holds), that is for the smoothed fields. However, even if we
were provided by Nature with an accurate 3D velocity field (as
accessible in N-body simulations) we would still prefer to smooth it,
in order to reduce the nonlinearities. We would do so because linear or
weakly nonlinear density and velocity fields are strongly correlated
and the form of this correlation offers us a possibility to test the
gravitational instability hypothesis and to measure the value of
$\Omega$.

Smoothing of the fields is realised by averaging them with a certain
window of a certain scale $R$, $W_R$. For example, the smoothed
density contrast field is

\be 
\ov{\de}(\bfx) = \int {\rm d}^{3} x' \, \de(\bfx') \, 
W_R(\bfx -\bfx')  
\,.
\label{eq-57} 
\ee
The velocity fields are commonly smoothed with a Gaussian filter. In
this case, the second-order density--velocity divergence relation
still has the form~(\ref{eq-52}), but the coefficient $a_2$ is a
function of the power spectrum index $n$,
\be    
a_2 = \f{1 + K -2C}{3} \ _{2} F_{1} \left( \f{n+3}{2}, \f{n+3}{2},
      \f{5}{2}, \f{1}{4} \right) 
\,.
\label{eq-58} 
\ee
Here, $_{2} F_{1}$ is the hypergeometric function (C{\L}97). The effective
power-law index $n$ is the slope of the $\log{P}$--$\log{k}$ relation
at the smoothing scale $R$ \cite{b94}. For $n = -3$ the
coefficient $a_2$ is $(1 + K -2C)/3$, equal to that for the case of no
smoothing, equation~(\ref{eq-53}). The factor $(1 + K -2C)/3$ can be
well approximated by its value for $\Omega = 1$: $4/21 \simeq
0.19$. For higher spectral indices $a_2$ grows monotonically up to the
value $\simeq 0.30$ for $n = 1$. At the scales of interest, i.e.\ of
several megaparsecs, the effective index of the {\em observed} power
spectrum is clearly different from the value $n = - 3$; e.g.\ for
\textit{IRAS} galaxies it is $n = -1.4$ \cite{fish}. Also, the
power spectra which are commonly used in N-body simulations have the
values of the effective index different from $n = - 3$; e.g.\ for the
standard CDM, for $R > 5 \hmpc$, $n \ge -1$ (e.g.\ C{\L}97). Therefore,
equation~(\ref{eq-52}), when applied to the smoothed fields,
underestimates density because the value of the coefficient $a_2$ is
underestimated. Specifically,
\be
\calD = \Delta a_2 ( \de^2 - \veps^2 ) + \calO(\veps^3)
\,,
\label{eq-59} 
\ee
where $\Delta a_2 \equiv a_2(-3) - a_2(n)$. The residual $\calD$ is
thus a parabola in $\de$ with a negative coefficient. Finally,
relation~(\ref{eq-26}), equivalent to the Gramann solution, must yield
the same residual, since $\de$-$\varte$ relation~(\ref{eq-52}) is its
version averaged over the shear. This is indeed observed in the
simulations: the Gramann solution yields essentially the same
parabolic residual as the $\de$-$\varte$ approximation of Bernardeau
(1992), derived also for the case of no smoothing.

How to extend equation~(\ref{eq-26}), or its another
form~(\ref{eq-41}), for the case of smoothed fields?  Applying a 
smoothing filter to both sides of equation~(\ref{eq-41}) we have
\begin{eqnarray} 
\ov{\de}(\bfx) 
&=& \ov{\varte}(\bfx) + \f{1 + K -2C}{3} \ov{\varte^2}(\bfx) 
- \f{1 + K -2C}{2} \ov{\Sig^2}(\bfx)  
\nonumber \\
&+& \calO(\veps^3)
\,,
\label{eq-60} 
\end{eqnarray}
where e.g.\ the smoothed $\ov{\varte^2}$ is given by
\be 
\ov{\varte^2}(\bfx) = \int {\rm d}^{3} x' \, \varte^2(\bfx') \, 
W_R(\bfx -\bfx')  
\,.
\label{eq-61} 
\ee
However, from observations we can estimate only smoothed fields, and
only after smoothing we can perform transformations on them (like
squaring). Our purpose is therefore to express the right-hand side of
equation~(\ref{eq-60}) as a function of $\ov{\varte}$ and

\be
\ov{\Sig} \equiv 
\left( \ov{\Sig}_{ij} \ov{\Sig}_{ij} \right)^{1/2}
\,.
\label{eq-62} 
\ee
Since smoothing and nonlinear transformations do not commute, in
general $\ov{\varte^2}$ is not equal to $\ov{\varte}^2$; similarly for
the shear. It is most clearly seen by writing
\be 
\ov{\varte^2} = \ov{(\varte - \ov{\varte})^2} + \ov{\varte}^2 \ge 
\ov{\varte}^2
\,,
\label{eq-62a} 
\ee
wherefrom $\ov{\varte^2}$ is equal to $\ov{\varte}^2$ only when the
spectral index $n = -3$, i.e.\ when the fluctuations have so large
wavelengths that $\ov{\varte} = \varte$. Moreover, the same values of
$\ov{\varte}$ can lead to different values of $\ov{\varte^2}$. (As a
simplest academic example the reader can consider one-dimensional
fields $\varte_1 = 1$ and $\varte_2 = 2 x$, smoothed with a top-hat
filter over the segment $[0,1]$.) It means that the relation between
$\ov{\de}(\bfx)$ and the variables $\ov{\varte}(\bfx)$ and
$\ov{\Sig}(\bfx)$ is non-deterministic. Again, the mean trend is given
by the conditional mean,

\be 
\left. \left\lan \ov{\de} \right\ran 
\right|_{\ov{\varte},\ov{\Sig}} 
= \ov{\varte} + \f{1 + K -2C}{3}
\left. \left\lan \ov{\varte^2} - \f{3}{2} \ov{\Sig^2} \right\ran
\right|_{\ov{\varte},\ov{\Sig}} 
\,.
\label{eq-63} 
\ee

The standard approach would be to derive the joint PDF for the four
variables $\ov{\varte^2}$, $\ov{\Sig^2}$,
$\ov{\varte}$, and $\ov{\Sig}$,
$p(\ov{\varte^2},\ov{\Sig^2},\ov{\varte},
\ov{\Sig}$), and to integrate over it the second term on the
right-hand side of the above equation. Fortunately, this horrible
calculation is unnecessary because the result can be simply
guessed. Firstly, it must be a quadratic form in $\ov{\varte}$ and
$\ov{\Sig}$, since it comes from second-order perturbative
contributions. Second, when averaged over $\ov{\Sig}$, it must
reduce to the second term of the smoothed version of
equation~(\ref{eq-52}),
\be
\left. \left\lan \ov{\de} \right\ran \right|_{\ov{\varte}} 
= \ov{\varte} + a_2 ( \ov{\varte}^2 - \ov{\veps}^2 ) 
\,,
\label{eq-64} 
\ee
where $a_2$ is given by equation~(\ref{eq-58}) and $\ov{\veps}^2
\equiv \left\lan \ov{\de}^2 \right\ran$. 

I postulate that 
\be 
\f{1 + K -2C}{3}
\left. \left\lan \ov{\varte^2} - \f{3}{2} \ov{\Sig^2} \right\ran
\right|_{\ov{\varte},\ov{\Sig}} = 
a_2 \left( \ov{\varte}^2 - \f{3}{2} \ov{\Sig}^2 \right)
\,.
\label{eq-65} 
\ee
The first condition is obviously satisfied. Similarly to
equation~(\ref{eq-47}), we have
\begin{eqnarray}
\left\lan \ov{\varte} \, \ov{\Sig}_{ij} \right\ran 
&=& \int \f{{\rm d}^3 k}{(2 \pi)^3} \, 
\left( \hat{k_i} \hat{k_j} - \f{1}{3} \de_{ij} \right)
P(k) W^2_R(k)
\nonumber \\
&=& 0
\,.
\label{eq-66} 
\end{eqnarray}

Here, $W_R(k)$ is the Fourier transform of the window function. Thus,
also the smoothed fields $\ov{\varte}$ and $\ov{\Sig}$ are
statistically independent. It implies that $\lan \ov{\Sig}^2
\ran|_{\ov{\varte}} = \lan \ov{\Sig}^2 \ran = (2/3) \ov{\veps}^2$
(eq.~[\ref{eq-51}]), hence the second condition is satisfied as
well. Finally, an additional term $\propto \ov{\varte}\, \ov{\Sig}$ on
the right-hand-side of equation~(\ref{eq-65}) would violate it,
because $\lan \ov{\varte}\, \ov{\Sig} \ran|_{\ov{\varte}} =
\ov{\varte} \lan \ov{\Sig} \ran \ne 0$: the average of the
shear scalar does not, unlike the average of its component
$\ov{\Sig}_{ij}$, vanish since $\ov{\Sig}$ is positive-definite.

Thus, the form on the right-hand-side of equation~(\ref{eq-65}) is
the unique quadratic form in $\ov{\varte}$ and $\ov{\Sig}$ which, when
averaged over $\ov{\Sig}$, reduces to the second term of
equation~(\ref{eq-64}). Therefore, postulated equation~(\ref{eq-65})
is indeed correct. Combining it with~(\ref{eq-63}) we obtain
\be 
\left. \left\lan \ov{\de} \right\ran \right|_{\ov{\varte},\ov{\Sig}} 
= \ov{\varte} + a_2 \left( \ov{\varte}^2 - \f{3}{2} \ov{\Sig}^2 \right)
\,,
\label{eq-67} 
\ee
or, using equations~(\ref{eq-21}) and~(\ref{eq-42}), 
\be 
\left. \left\lan \ov{\de} \right\ran \right|_{\ov{\te},\ov{\s}} 
= - f^{-1} \ov{\te} + a_2 f^{-2} 
\left( \ov{\te}^2 - \f{3}{2} \ov{\s}^2 \right)
\,.
\label{eq-68} 
\ee
Here, $\ov{\te}$ and $\ov{\s}$ is the expansion and shear of the
{\em smoothed} velocity field $\ov{\bfv}$, 
\be
\ov{\te} \equiv \nabla_{\bfx} \cdot \ov{\bfv}
\label{eq-69} 
\ee
and
\be
\ov{\s} \equiv 
\left( \ov{\s}_{ij} \ov{\s}_{ij} \right)^{1/2}
\label{eq-70} 
\ee
with
\be
\ov{\s}_{ij} \equiv \f{1}{2} \left( \ov{v}_{i,j} +
\ov{v}_{j,i} \right) - \f{1}{3} \ov{\te} \de_{ij}
\,.
\label{eq-71} 
\ee

Equation~(\ref{eq-67}), with the coefficient $a_2$ given by
equation~(\ref{eq-58}), constitutes an extension of
equation~(\ref{eq-41}) for the case of smoothed fields. As already
discussed, smoothing of the fields induces spread in the relation
between the smoothed density and the smoothed expansion and shear in a
given point. In the present paper I will not attempt to compute the
spread. However, it is certainly smaller than the spread in the
smoothed density--velocity divergence relation~(\ref{eq-64}), since
this relation is obtained from~(\ref{eq-67}) by averaging over the
shear. This averaging is an extra source of the spread in the
$\de$--$\varte$ relation: for unsmoothed fields the spread is given by
equation~(\ref{eq-56}), while the relation~(\ref{eq-41}) between
density and the two velocity scalars is entirely deterministic.

It should be stressed that equation~(\ref{eq-67}) assumes Gaussian
initial conditions, since the coefficient $a_2$,
equation~(\ref{eq-58}), has been computed by C{\L}97 under this
assumption. For non-Gaussian initial conditions, a detailed form of
the relation between the smoothed density and the smoothed expansion
and shear remains to be studied. 

\section{Comparison to N-body simulations} 

Ganon \etal (1997) have performed N-body simulations for a CDM family
of models. For the fields smoothed with a Gaussian window of radius $R
= 12 \hmpc$, Ganon \etal have estimated the value of the coefficient 
$a_2$ to be

\be
a_2^{(NB)} \simeq 0.28
\label{eq-72} 
\ee
(cf.\ also Willick \etal 1997). As explained in the previous section,
the Gramann approximation yields 

\be
a_2^{(G)} = \f{4}{21} \simeq 0.19 \,.
\label{eq-73} 
\ee

What is the value of $a_2$ predicted by the approximation derived
here? The hypergeometric function in expression~(\ref{eq-58}) can be
expanded in powers of $n+3$, simply by rearranging the appropriate
terms of the hypergeometric series (cf.~{\L}okas \etal 1995). In the
expansion, the term linear in $n+3$ vanishes. The resulting formula
for the coefficient $a_2$, truncated at the cubic term, is

\be
a_2 = \f{4}{21} \left[ 1 + c_2 (n+3)^2 + c_3 (n+3)^3 \right]
\,, 
\label{eq-74} 
\ee
where 
\be
c_2 = 0.02139 \qquad \hbox{and} \qquad c_3 = 0.00370
\,.
\label{eq-75} 
\ee
The polynomial~(\ref{eq-74}) approximates expression~(\ref{eq-58}) in
the range $-3 \le n \le 1$ with accuracy of $0.2$\% or better. If
$\Omega \ne 1$, the factor $4/21$ should be replaced by the factor $(1
+ K -2C)/3$, weakly varying with $\Omega$.

For a smoothing radius of $12 \hmpc$, the effective index of the
standard CDM spectrum is (for details see C{\L}97)

\be
n = -0.404
\,.
\label{eq-76} 
\ee
Using formula~(\ref{eq-74}), the predicted value of the coefficient
$a_2$ is then

\be
a_2 \simeq 0.23
\,.
\label{eq-77} 
\ee This value lies half-way between $a_2^{(G)}$ of the Gramann
approximation, equation~(\ref{eq-73}), and $a_2^{(NB)}$ estimated from
N-body by Ganon \etal (1997), equation~(\ref{eq-72}). It means, on the
one hand, that the density--velocity relation derived in this paper,
equation~(\ref{eq-68}) with $a_2$ given by equation~(\ref{eq-58})
or~(\ref{eq-74}), is not in full agreement with the results of N-body
by Ganon et al.\@ On the plot of the difference between the approximate
and the true density as a function of the true density, inferred from
N-body, it will still leave some parabolic residuals. On the other
hand, however, these residuals will be smaller, roughly two times,
than for the Gramann approximation. As stated earlier, the Gramann
approximation is valid only for unsmoothed fields. Not surprisingly
then, the proper second-order formula for smoothed fields is a better
estimator of smoothed density from smoothed velocity.

The slight discrepancy between the value~(\ref{eq-77}) and that
estimated from N-body can be attributed to perturbative contributions
of higher orders. Here, or in the paper of C{\L}97, the value of $a_2$ has
been derived at the lowest relevant order in perturbation theory, the
second. Higher-order corrections yield a contribution to $a_2$ which
is of the order of $\ov{\veps}^2 = \lan \ov{\de}^2 \ran$. In other
words,

\be
a_2(\ov{\veps}) = a_2(\ov{\veps} \to 0) + p_{} \ov{\veps}^2 + 
\calO(\ov{\veps}^4)
\,,
\label{eq-78} 
\ee
where the dimensionless coefficient $p_{}$ remains to be
calculated. Since the (linear) variance of the field in question
(Gaussian smoothed with radius of $12 \hmpc$) is $\ov{\veps}^2 =
0.076$, the value $p_{} \sim 0.7$ is sufficient to account for the
discrepancy between $a_2$ and $a_2^{(NB)}$.

On the other hand, it is important to confirm the results of Ganon
\etal (1997) by independent simulations. Preliminary results of such
show that the coefficient $a_2$ has indeed slightly higher value than
predicted by second order perturbation theory, approaching it for
small $\ov{\veps}$ \cite{bcl,chod}.

\section{Summary}
In the present paper I have studied the relation between the weakly
nonlinear cosmic density and velocity fields. I have derived up to
second order in perturbation theory an expression for density as a
purely local function of the expansion (divergence) and shear of the
velocity field (eq.~[\ref{eq-26}] or~[\ref{eq-41}]). The relation
depends on $\Omega$ both strongly and weakly. I have shown that the
Gramann~(1993) solution (eq.~[\ref{eq-29}]) is equivalent to
equation~(\ref{eq-26}) with the weak $\Omega$-dependence
neglected. Also, a similar relation has been independently found by
Catelan \etal (1995).

The locality of the relation derived here is in contrast with the
non-locality of the density--velocity divergence relation calculated
by C{\L}97. I have shown that averaging of equation~(\ref{eq-41}) over
shear given divergence yields up to second order the formula of C{\L}97
(eq.~[\ref{eq-52}]); thus, the source of the spread in the latter
relation is the distribution of shear. Since the form of this
distribution is known, it is straightforward to compute the
higher-order conditional moments of the density--velocity divergence
relation. In particular, I have computed the spread of this relation
(the conditional variance; eq.~[\ref{eq-56}]) and have found it to
coincide with the result of rather cumbersome calculations by
Chodorowski, {\L}okas \& Pollo (1997).

Smoothing is a necessary ingredient of large-scale density--velocity
comparisons. I have then generalized equation~([\ref{eq-26}]) for the
case of smoothed fields (eq.~[\ref{eq-68}] with $a_2$ given by
eq.~[\ref{eq-58}]). Smoothing not only modifies the shape of the
relation between density and the two velocity scalars but also induces
the spread in it. I have not computed the spread explicitly. Still, I
have shown that it is smaller than the corresponding spread in the
density--velocity divergence relation, since that spread has an extra
source: averaging over shear given divergence.

Equation~(\ref{eq-26}) for unsmoothed fields does not depend on the type
of initial conditions. Its smoothed counterpart~(\ref{eq-68}), however,
assumes Gaussian initial conditions; a detailed form of it for
non-Gaussian models remains to be investigated.

Checks against N-body simulations show that relation~(\ref{eq-68}) is
still a slightly biased estimator of large-scale density from velocity
(section~6). However, it was not the main concern of this paper. The
ultimate goal of attempts to establish semi-linear relations between
the density and velocity fields is to construct a local estimator of
density from velocity which is not only unbiased but has minimum
variance as well. The present paper is the first attempt to address
the second point. As stated earlier, the inclusion of the shear term
in the relation between the large-scale density and velocity reduces
its spread. In this paper I have demonstrated how to include the shear
in the second-order perturbative relation. In future, I plan to do so
for higher-order relations as well.

The relation density versus velocity expansion and shear,
derived here, is a new, lower-variance estimator of density from the
velocity field that can be applied in the cosmic density--velocity
comparisons.

\section*{Acknowledgments}

I am grateful to Roman Juszkiewicz for numerous stimulating
discussions. I wish to thank Francis Bernardeau, Ewa {\L}okas and Adi
Nusser for useful suggestions on a preliminary version of this
paper. An anonymous referee is warmly acknowledged for valuable
comments which helped to improve the presentation of the results. I
acknowledge also hospitality of Fran{\c c}ois Bouchet and Alain Omont
at Institut d'Astrophysique de Paris. This research has been supported
in part by the Polish State Committee for Scientific Research grant
No.~2P30401607, the French Ministry of Research and Technology within
the program PICS/CNRS No.~198 (Astronomie Pologne) and Human Capital
and Mobility program of the European Communities (ANTARES).

\bsp

\end{document}